\documentclass[3p,amsmath,amssymb,compress]{elsarticle}
\usepackage{amsmath}
\usepackage{graphicx}
\usepackage{latexsym}
\usepackage{amsmath,amssymb}
\usepackage{bm} 
\usepackage{xcolor}
\usepackage{stackrel}
\usepackage{accents}
\usepackage{braket}

\journal{Annals of Physics: contribution to the Philip W. Anderson Memorial Special Issue}

\numberwithin{equation}{section}

\newcommand{\KC}{\overset{\circ}{,}}
\renewcommand{\v}[1]{\textbf{\textit #1}}

\begin{document}
\begin{frontmatter}	

\title{Quantum kinetics of anomalous and nonlinear Hall effects \\ in topological semimetals}
\author[MPI]{Elio J. K\"onig}
\author[UW]{Alex Levchenko}
\address[MPI]{Max Planck Institute for Solid State Research, Heisenbergstrasse 1, D-70569 Stuttgart, Germany}
\address[UW]{Department of Physics, University of Wisconsin-Madison, Madison, Wisconsin 53706, USA}

\begin{abstract}
We present a systematic microscopic derivation of the semiclassical Boltzmann equation for band structures with the finite Berry curvature based on Keldysh technique of nonequilibrium systems. In the analysis, an ac electrical driving field is kept up to quadratic order, and both cases of small and large frequencies corresponding to intra- and interband transitions are considered. In particular, this formulation is suitable for the study of nonlinear Hall effect and photogalvanic phenomena. The role of impurity scattering is carefully addressed. Specifically, in addition to previously studied side-jump and skew-scattering processes, quantum interference diffractive contributions are now explicitly incorporated within the developed framework. This theory is applied to multifold fermions in topological semimetals, for which the generic formula for the skew scattering rate from the Pancharatnam phase is obtained along with the corresponding anomalous Hall conductivity.  
\end{abstract}


\end{frontmatter}

\tableofcontents

\section{Introduction} 
\subsection{A brief historical tour} 

The anomalous Hall effect (AHE), including its spin, valley, thermoelectric, and quantized counterparts, in their complexity represent perhaps the most fascinating set of quantum transport phenomena, see Refs.~\cite{Sinova-AHE-RMP,Niu-RMP,Hughes-ARCMP,Sinova-SHE-RMP,SCZ-Qi-ARCMP,Niu-RPP} for reviews and references herein. At the same time, anomalous transport coefficients provide key insights to correlated and topological quantum materials. This includes the intriguing normal state of cuprate superconductors, with an anomalous temperature scaling of Hall and longitudinal resistivity which Anderson traced back to his theory of tomographic Luttinger liquids~\cite{Anderson1991,OngAnderson1991}, as well as a giant enhancement of the thermal Hall effect~\cite{ZhangHardy2001}. In the presence of spin-orbit coupling, a large anomalous Hall effect also gives access to the physics of local moments, first discussed by Kondo~\cite{Kondo1962} and later extended to the mixed valence regime by Coleman, Anderson and Ramakrishnan~\cite{ColemanRamakrishnan1985,RamakrishnanAnderson1985}, who demonstrate a massively enhanced skew scattering off quantum impurities. Finally, Kerr rotation experiments~\cite{Kapitulnik2009}, which probe the finite frequency anomalous Hall conductance, have become one of two litmus tests for time reversal symmetry breaking superconductors, i.e. solid state cousins of unconventional pairing states pioneered by Anderson, Brinkman and Morel in the context of $^3$He~\cite{AndersonMorel1961,BrinkmanAnderson1973} (the other litmus test being muon-spin-rotation).

In retrospect, when counting from the experimental discovery by Edwin Hall in 1880 it took seven decades for
a comprehensive theory of the AHE to be developed even for weakly coupled materials. In their seminal work Karplus and Luttinger~\cite{KL-PRB1954} laid out foundations of the AHE by elaborating rigorous microscopic linear-response  calculations in multi-band metals, recognizing importance of spin-orbit interaction, and most importantly discovering an anomalous group velocity term in the semiclassical equation of motion of Bloch electrons in solids. During the following several decades the extrinsic mechanisms of anomalous Hall transport where uncovered and scrutinized. Smit~\cite{Smit-I,Smit-II} found a skew scattering mechanism of AHE that captures left-right asymmetry in the differential scattering cross section of a conduction electron on an impurity potential. Berger~\cite{Berger} studied another subtle quantum side-jump contribution, which occurs due to coordinate shifts in electron trajectories following the consecutive scattering events. 

The spin version of the Hall effect was proposed by Dyakonov and Perel~\cite{Dyakonov-Perel-I,Dyakonov-Perel-II}, which consists of spin accumulation by passing an electrical current. This insightful work ultimately triggered another cycle of numerous studies where intrinsic and extrinsic origins of the spin Hall effect (SHE) were debated. 

The discovery of the integer Hall effect by von Klitzing, Dorda, and Pepper~\cite{Klitzing-PRL1980} and its further explanation eventually greatly contributed to the deeper understanding of the AHE through the work of Thouless, Kohmoto, Nightingale, and den Nijs (TKNN)~\cite{TKNN} who derived their famous invariant for the Hall conductance formula. The next crucial steps was done by Haldane~\cite{Haldane} who presented a concrete model for the quantized anomalous Hall effect (QAHE) in a lattice system. Shortly after that, the geometric and topological aspects of QAHE were understood and quickly became the dominant paradigm. The anomalous velocity was connected to the Berry phase~\cite{Berry} so it became possible to express the anomalous Hall conductance in terms of the integral of the Berry curvature over the momentum space~\cite{Chang-Niu-PRL95}. This ultimately gave a representation of quantization in terms of the Chern number of fully filled bands. 

These advances elucidated the geometrical origin of the AHE even in the situations when it is not quantized. This is in fact true not only for the intrinsic mechanism, but also applicable to extrinsic mechanisms: skew scattering amplitudes as well since coordinate shifts in the side-jump processes can be expressed in terms of Pancharatnam phase~\cite{Pancharatnam}, which represents a special case of Berry phase. Indeed, gauge invariant formulae for coordinate displacements of electrons undergoing quantum transitions were known early on from work of Belinicher, Ivchenko, and Sturman~\cite{BelinicherSturman1982} on the kinetic theory of photovoltaic effect, however, the topological aspects of the problem were not realized at that time. This work was rediscovered only relatively recently and provided renewed important insights~\cite{SinitsynMacDonald2006}. The peculiar observation is that at least in some models of the disorder potential the microscopic characteristics of the impurity potential drop out from the final expression for the coordinate shifts that thus can be expressed only in terms of electronic Bloch functions. This feature makes certain extrinsic contributions to some extent universal.   

The quantum version of the spin Hall effect (QSHE) was discussed by Kane and Mele~\cite{Kane-Mele} in the context of a hexagonal lattice model as in graphene. The description was constructed from the generalized model of Haldane extended to spin-$\frac{1}{2}$ electrons with spin-orbit coupling. However, weak spin-orbit interaction in graphene stimulated further searches of material platforms where this effect could manifest. The CdTe/HgTe heterostructure was proposed by Bernevig, Hughes, and Zhang~\cite{BHZ-QSH} to have the right ingredients with an inverted band structure for the proper thickness of quantum wells when a topological phase transition occurs. This prediction was confirmed experimentally~\cite{Konig-QSH} via observed robust conductance quantization facilited by topologically protected edge modes. The time reversal broken analog, i.e. the quantum anomalous Hall effect was recently observed in thin films of magnetic topological insulators~\cite{QAHE-TI} as well as quantum valley Hall effect in graphene superlattices~\cite{QVHE-Graphene}. More generally, graphene bilayers at magic twist angle provide the most recent experimental platform for the study of interaction driven emergent ferromagnetic state exhibiting giant anomalous Hall response~\cite{TBLG-AHE} and possibly topological Chern insulating behavior. 

From the present day perspective, it is perhaps fair to say that it took the physics community a whole century to fully understand and conceptualize the plethora of anomalous Hall effects. This is certainly a noble achievement, but in hindsight one is left with a sobering realization that all that intellectual advance concerns essentially a single particle physics in the linear response transport problem. To a large extent we still know very little about effects of interactions~\cite{Langenfeld-Wolfle,Muttalib-Wolfle,Badalyan-Vignale,SLAL,Avdoshkin-Kozii,Ipsitamandal2020}, quantum interference corrections~\cite{Dugaev-AHE-WL,Muttalib-AHE-WL,Meier-AHE-WL,KoenigMirlin2014}, and strong nonequilibrium conditions on linear and nonlinear anomalous Hall responses. In part this motivates our work to advance current theory further and bridge the gap towards modeling of experiments. 

\subsection{Technical synopsis of previous works}

At the technical level there exist several distinct approaches to describe generalities of the anomalous Hall effect. An analytical theory based on a transport like equation for the density matrix was developed by Luttinger~\cite{Luttinger}. In this framework effects of impurities can be incorporated perturbatively by a systematic expansion in disorder potential. This treatment provides rigorous and controllable quantum-mechanical description which also reveals importance of the off-diagonal elements of the density matrix and of the velocity operator in the calculation of conductivity tensor. However, this method is not very practical and difficult to implement even in relatively simple model cases. 

A precursor to modern semiclassical theory is the work by Adams and Blount~\cite{Adams-Blount,Blount1962} who used the picture of particle wave-packet dynamics in a crystal field described by noncommuting coordinates. In a series of applications this formulation was generalized to degenerate bands in the context of group III-V n-type semiconductors~\cite{Chazalviel,Nozieres-Lewiner,Lyo-Holstein}, which initially had difficulties with incorporation of extrinsic contributions. A modern version of these theories can be presented in the form of Eilenberger-type equations for the reduced Green's function, which incorporates both short-range impurity scattering, non-Abelian Berry curvature terms stemming from the band degeneracies, as well as quantum anomalies ~\cite{Shelankov,WongTserkovnyak2011,WicklesBelzig2013,Bolet2014,SekineMacDonald2017}. 

A fully semiclassical description based on the Boltzmann equation, a diagrammatic method based on the Kubo-Streda formulas, as well as Keldysh technique, were applied to the problem of AHE, see Ref.~\cite{Sinitsyn-JPCM} for an overview. It should be stressed that establishing a connection between these methods is not merely a trivial exercise and requires quite laborious calculations~\cite{SinitsynSinova2007,KovalevSinova2008}. For instance, at the level of the Boltzmann equation, one finds an additional contribution to transverse conductivity termed as an anomalous distribution. Indeed, the emergent asymmetry of the distribution function, even without an asymmetry in the kernel of collision integral, is the result of the side-jump process in the presence of external electric field which modifies energy conservation. The corresponding correction to the distribution function combined with the conventional part of the band velocity leads to the additional Hall current. This term is not immediately evident at the level of diagrammatic Green's function calculation~\cite{Mitscherling2020} as it is absorbed into the part of side-jump contribution. This also suggests that clear separation of various terms is somewhat ambiguous. At the same time, diagrammatics produces new terms that were missed in all previous approaches. These are hybrid~\cite{KovalevSinova2008} and diffractive~\cite{AdoTitov2015,KoenigLevchenko2016,EKAL-PRL17} skew scatterings. The former one is inversely proportional to the impurity concentration, thus resembling the usual skew scattering from non-Gaussian disorder, but it is independent of impurity strength, resembling the side-jump mechanism. The diffractive process is present already at the level of Gaussian disorder and, perhaps counterintuitively, is independent of both impurity concentration and impurity strength due to subtle cancellations, so it scales as the intrinsic term. The validity of these results has been established in different models~\cite{Milletari-Ferreira-PRB16,Ado-PRL16,EKAL-PRL17,Ado-PRB17}. 

\subsection{Recent developments: nonlinear anomalous responses}

The first systematic attempt to extend the semiclassical theory of AHE to the domain of nonlinear Hall responses was presented in the paper by Deyo \textit{et al}.~\cite{DeyoSpivak2009}. The emphasis of the study was put on the linear and circular photogalvanic phenomena, including calculation of the corresponding response tensors in the presence of weak static magnetic field. The analysis was carried out for the bulk crystal symmetries $T_d$ and $C_{6v}$, and the point symmetry $C_s$ of a quantum-well structures. Up to that point such calculations were rigorously established only in the clean limit of semiconductor structures~\cite{SipeShkrebtii2000}. In part motivated by experiments~\cite{Orenstein-NP17,Gedik-NP17} further extensions of the theory and applications were tailored towards circular photogalvanic effect in Weyl semimetals~\cite{deJuanMoore2017,Morimoto-PRB16,KoenigLevchenko2017,Golub-Ivchenko}, which can be considered as an ac non-linear Hall response at optical frequencies.

In a parallel line of developments, Sodemann and Fu~\cite{SodemannFu2015} demonstrated the topological origin of the transverse Hall-like currents that occur in second-order response to an external electric field. Arguments were put forward that these effects can occur in a wide class of two- and three-dimensional time-reversal invariant and inversion breaking materials, including topological crystalline insulators, transition metal dichalcogenides and Weyl semimetals~\cite{SinghVanderbilt2020}. The crucial distinction from the linear AHE is that these nonlinear response functions are governed by the dipole moment of the Berry curvature in momentum space. Furthermore, the Berry curvature dipole emerges both in the dc current and also in the second harmonic. These initial results were extended to include effects of skew-scattering and side-jump on the second order responses~\cite{KoenigPesin2019,IsobeFu-SA19,DuXie-NC19,NandySodemann-PRB19,XiaoNiu2019}. 

\subsection{Overview of this work}

In this work we construct kinetic theory of quantum transport in multiband materials with nontrivial band topology. We base our analysis on Keldysh technique for nonequilibrium systems. In the analysis, external potentials are kept to quadratic order which enables us to address nonlinear anomalous transport effects. This includes both intraband processes at low frequency and interband photogalvanic responses. A careful attention is paid to impurity scattering effects as we retain in calculations disorder potential up to the fourth order. This automatically includes all known skew scattering and side jump disorder-induced contributions, and also less studied quantum interference processes. Even though we employ formal machinery to derive kinetic equation, we use semiclassical language and interpretation of emergent terms. For instance at the level of the Dyson equation for self-energies we make a connection between diagrammatic and semiclassical approaches. To avoid spurious difficulties at the intermediate steps, we use fully gauge-invariant construction and incorporate Berry connection explicitly in the Wigner transform of operators. Given already quite a laborious task, we do not consider effects of external magnetic field.       

\subsection{Multifold fermions}

As an application of our theory, in Sec. \ref{sec:Multifold} we present the first microscopic study of anomalous impurity scattering in models of multifold fermions, which are generalizations of 2D and 3D Weyl and Dirac semimetals~\cite{Manes2012,WiederKane2016,BradlynBernevig2016,BouhonBlackSchaffer2017,TangZhang2017} that contain $ (2S + 1)-$fold degenerate touching point. We also use the terminology of multifold fermions, when the degeneracy at touching point is lifted (in the simplest case, this lifting occurs by a mass gap). The kinetic part of the Hamiltonian in the simplest $\v k \cdot \v p$ expansion, is of the form
\begin{equation}
    H_{\rm kin}(\v p) = d_0(\v p) + \sum_{i=1}^3 d_i(\v p) S_i. \label{eq:Multifold}
\end{equation}
Here, $d_{0,1,2,3}(\v p)$ are momentum dependent functions and $S_i$ are spin-$S$ matrices, with the usual commutation algebra $[S_i,S_j]=i\varepsilon_{ijk}S_k$, where $\varepsilon_{ijk}$ is the Levi-Civita symbol. The Hamiltonian~\eqref{eq:Multifold} thereby generalizes the familiar $S = 1/2$ case applicable, e.g., to Weyl semimetals in 3D and gapped topological insulator surface states in 2D. Particularly 3D multifold fermions~\cite{SchroeterChen2019} have been of great interest for anomalous transport and optics~\cite{FlickerGrushin2018,SanchezMartinezGrushin2019,Habe2019,NandyRoy2019,SadhukhanNag2020,MaulanaPronin2020}, which is related to the fact that RhSi and CoSi contain 4- and 6-fold fermionic touching points~\cite{ChangHasan2017} and display signatures of quantized photocurrent generation~\cite{deJuanMoore2017,KoenigLevchenko2017,ReesOrenstein2020}. At the same time, 2D multifold fermions may appear as topological surface states~\cite{WiederBernevig2018} or in appropriately designed lattice models~\cite{BerciouxHausler2009,GreenChamon2010,ApajaManninen2010,ShenXing2010,Essafi2017,ZhuZhu2017,WangYao2018}.

In our study we present general results for $S = 1$ and $S = 3/2$, and evaluate the impurity scattering effects in the case of $S=1$ fermions in detail. We contrast these results to the most well studied $S=1/2$ situation in the model of massive 2D Dirac fermions. 

\subsection{Outline of the paper}

The rest of the paper is organized as follows. In Sec. \ref{sec:BKE} the general setup is introduced, along with assumptions and notations. After defining the modified Wigner transform and Moyal expansion, the Dyson equation for the Keldysh block of the Green's function is discussed in details. This block has a non-diagonal structure in the band index and we outline the systematic solution strategy for the off-diagonal quantum components to arrive at kinetic equations for the band-diagonal distribution functions that have natural semiclassical interpretation. These equations are further simplified by following the usual gradient expansion. From the projections onto corresponding bands the collision terms are derived for various processes and topological terms are identified. In the next Sec.~\ref{sec:Current} the electrical current operator is considered and its semiclassical form is deduced with the side-jump contribution made explicit. These two sections represent the core of the paper. Finally, as mentioned, we present an application to multifold fermions in Sec.~\ref{sec:Multifold} and conclude with a summary and outlook in Sec. \ref{sec:Summary}. Some additional technicalities concerning collision kernels and momentum averages are presented in the appendix.   


\section{Derivation of quantum kinetic equation}\label{sec:BKE}

In this section we present a systematic derivation of the quantum kinetic equation in Berry curved matter in the presence of impurity scattering and external nonlinear drive. 

\subsection{Setup and Assumption}

We consider generic $N$-band Hamiltonian containing kinetic and potential terms
\begin{equation}
H = H_{\rm kin}(\v p) + U(t,\v x).
\end{equation}
We now discuss each of these two terms in detail.

\subsubsection{Kinetic energy and band structure}

\begin{figure}
\begin{center}
\includegraphics[width  = .75 \textwidth]{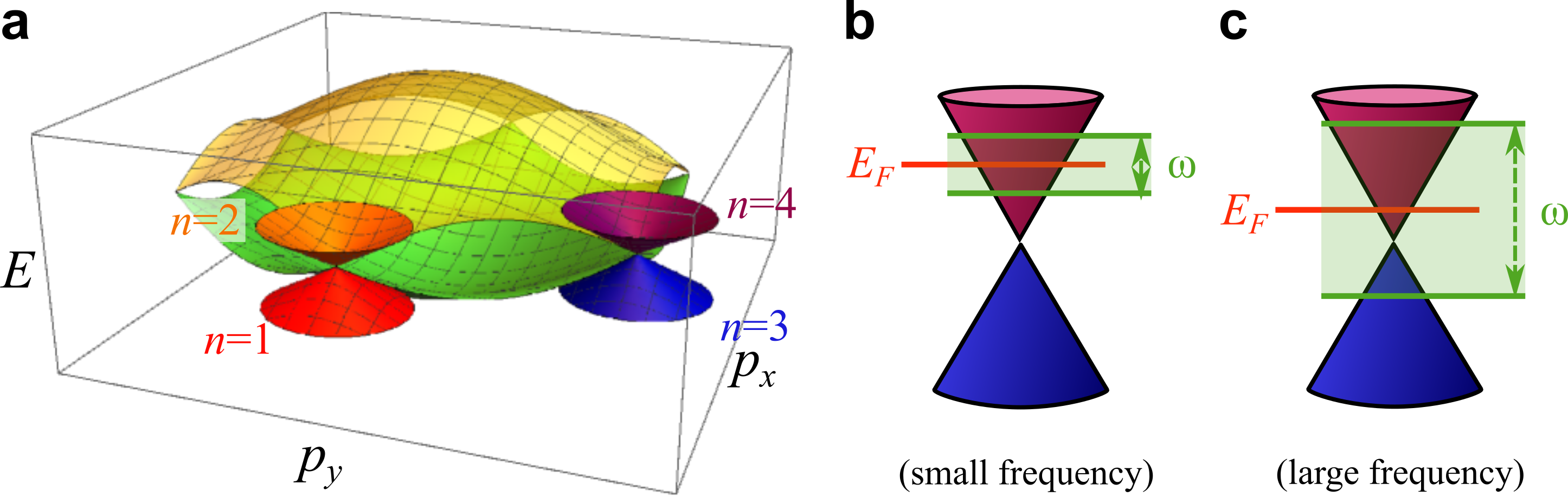}
\caption{Illustration of the convention. {\bf a} We use an $N$-band model (bands are labelled $n = 1, \dots, N$) based on the $\v k\cdot\v p$ expansion about special points (denoted by $\v b_{n}$) in the Brillouin zone. Here, we exemplify this for the tight binding model on a hexagonal lattice, where we keep only the states near $K$ and $K'$ nodes and thus $N = 4$, $\v b_1 = \v b_2 = (4\pi/3,0)/a$ and $\v b_3 = \v b_4 = (2\pi/3, 2\pi/\sqrt{3})/a$. {\bf b} ({\bf c}) Illustration of the possible small (large) frequency regime $\omega \ll E_F$ ($\omega \gtrsim E_F$).}
\label{fig:IllustrationNotation}
\end{center}
\end{figure}

We use an effective description of $H_{\rm kin}(\v p)$ by means of $\v k \cdot \v p$ Hamiltonians centered at $N_c$ (avoided) nodes in the Brillouin zone denoted $\v b_n$, see Fig.~\ref{fig:IllustrationNotation} for an illustration with $N = 4$ and $N_c = 2$ (note that $\v b_{1} = \v b_{2}$ and $\v b_{3} = \v b_{4}$ in this case). The solutions of the free Schr\"odinger equation are denoted as
\begin{equation}
\psi_{n, \v p}(\v x) = \braket{\v x \vert \psi_{n, \v p}} = e^{i (\v p - \v b_{n})\cdot \v x}\ket{u_{n, \v p}},     
\end{equation}
and we use the index $n = 1, \dots N$ to label the bands.

The $N-$vectors $\ket{u_{n, \v p}}$ constitute the sections of a fibre bundle over momentum space. In what follows we use the usual orthonormality and completeness relations $\langle u_{n, \v p}\vert u_{n',\v p}\rangle=\delta_{n,n'}$ and $\sum_n |u_{n, \v p}\rangle\langle u_{n, \v p}|=\boldsymbol{1}_N$. The gauge transformation $\ket{u_{n, \v p}} \rightarrow e^{i \phi_n(\v p)}\ket{u_{n, \v p}}$ is implied by the equivalence of solutions for local (in momentum space), $n$ dependent phase. We consider nondegenerate bands and define the Berry connection as 
\begin{equation}\label{eq:A}
\boldsymbol{\mathcal A}_{n n'}(\v p) = i \langle u_{n, \v p}\vert E \vert \nabla_{\v p} u_{n',\v p} \rangle,
\end{equation}
which is gauge covariant
\begin{equation}
\boldsymbol{\mathcal A}_{n n'} \rightarrow e^{-i \phi_n(\v p)}[\boldsymbol{\mathcal A}_{n n'} - \nabla_{\v p} \phi_n \delta_{n, n'}]e^{i \phi_{n'}(\v p)}.\label{eq:BerryConnectionTrafo}
\end{equation}
We have introduced the $N \times N$ matrix $E$ which is particularly simple if the dimension of the Hilbert space is equal at each of the $N_c$ nodes. Then it can be represented as a block matrix, within each block (diagonal blocks correspond to a given node) it is the $N/N_c$-dimensional identity,
\begin{equation}
E = \left (\begin{array}{ccc}
\mathbf 1_{N/N_c} & \cdots & \mathbf 1_{N/N_c} \\ 
\vdots &  \ddots & \vdots \\ 
\mathbf 1_{N/N_c} &  \cdots & \mathbf 1_{N/N_c}
\end{array} \right ).
\end{equation}
The Berry curvature of band $n$ is defined through the curl of connection vector from Eq. \eqref{eq:A}  
\begin{equation}\label{eq:BerryCurvature}
\boldsymbol{\Omega}_n = \nabla_{\v p} \times \boldsymbol{\mathcal A}_{nn}.
\end{equation} 

\subsubsection{Potential}

The scalar potential $U(t,\v x) = [\Phi (t, \v x) + V (\v x)] E$ contains a part $\Phi$ corresponding to the ac electric field and a part $V$ corresponding to the static disorder. The scalar potential may scatter between the $N_c$ nodes in momentum space, this is encoded in the $N \times N$ matrix $E$. 
We emphasize that this does not mean that internode scattering is always important: as the nodes are centered at different momenta, the off-diagonal elements acquire a phase factor. This leads to the finite momentum Fourier transform of the potential, which in turn is small for smooth $U(t,\v x)$. For a potential $U(t,\v x)$ to be able to scatter between nodes $n$ and $n'$, the Fourier component $U(t,\v q)$ must be sizable at $\v q = \v b_{n} - \v b_{n'}$.

For simplicity, we assume monochromatic ac field
\begin{equation}
\Phi(t,\v x) = \sum_{\xi = \pm} \Phi_\xi(\v x) e^{i \xi \omega t},\quad 
\Phi_\xi(\v x)=- e \boldsymbol{\mathcal E}_{\xi \omega} \v x, 
\end{equation}
where $\boldsymbol{\mathcal E}_{\xi \omega}  = \boldsymbol{\mathcal E}_{-\xi \omega} ^*$, and discuss the generalization to arbitrary $\Phi(t,\v x)$ in the results section. Throughout the derivation we will treat the following two cases, see Fig.~\ref{fig:IllustrationNotation} {\bf b}, {\bf c}.
In the first case, $\omega\ll E_F $, the frequency is small as compared to the Fermi energy $E_F$ i.e. all excitations are close to the Fermi surface.
In the second case, $\omega > E_F$, the frequency is large as compared to the Fermi energy and may lead to vertical interband transitions. We consider contributions up to second order in driving fields. 

The disorder potential is described by scalar impurities located at positions $\v R_i$ that we assume to be uniformly distributed in $\mathbb R^d$ with density $n_{\rm imp}$,
\begin{equation}
V(\v x ) = \sum_{i} \mathcal{V}(\v x-\v R_i).
\end{equation}
Our calculations are perturbative in powers of the impurity strength (see details in the next section).

\subsubsection{Wigner transform}

In addition to perturbatively weak impurity scattering, we assume an elastic scattering rate $1/\tau \propto n_{\rm imp} \vert \mathcal{V}(0)\vert^2 \nu(E)$, where $\nu(E)$ is the density of states), to be small as compared to the energy $E$ of the electrons. Then a semiclassical expansion is controlled.

The semiclassical phase-space quantization is conveniently expressed in terms of the Wigner transform 
\begin{align}
    {O}_{n n} (x,p) &= \int_{\Delta p} O_{nn}\left ( p+ \frac{\Delta p}{2},p- \frac{\Delta p}{2} \right)  e^{i \Delta p_\mu (x^\mu - \mathcal A^\mu_n(p))} \label{eq:GeneralizedWigner}
\end{align}
of band-diagonal matrix elements of an arbitrary operator $\hat O$
\begin{align}
{O}_{n, n'} (p; p') = \int_{tt'} e^{iEt - iE' t'}  \braket{\psi_{n, \v p} \vert   \hat{O} \vert \psi_{n', \v p'}} 
= \int_{xx'} \braket{u_{n, \v p} \vert   {O}(x,x') \vert u _{n', \v p'}} e^{- i x_\mu p^\mu + i x'_\mu {p'}^\mu} e^{i \v b_n \cdot \v x-i \v b_{n'} \cdot \v x} .
\end{align}
Here we introduced 1+d vectors $p = (E, \v p)$, $x = (t, \v x)$, $\mathcal A_n(p) =(0, \boldsymbol{\mathcal A}_{nn}(\v p))$ and used a Minkowski metric with signature $(-,+,\dots, +)$. Note that the Wigner transform defined in Eq.~\eqref{eq:GeneralizedWigner} differs from the conventional Wigner transform by the explicit account of Berry connection in the exponential~\cite{KoenigColeman2019}. This ensures that $O_{nn}(x,p) \rightarrow O_{nn}(x,p) + \mathcal O (\partial_x^2 \partial_p^2)$ is invariant under gauge transformations $\ket{u_{n, \v p}} \rightarrow e^{i \phi_{n, \v p}} \ket{u_{n, \v p}} $ up to subleading orders in the semiclassical gradient expansion. A similar idea was first used by Altshuler~\cite{Altshuler1978} to account for the gauge invariance in real space $\psi(\v x) \rightarrow e^{i \phi(\v x)} \psi(\v x)$ in the presence of an external electromagnetic vector potential. 

As usual, the Wigner transform of convoluted operators defines the Moyal expansion (we drop the index $nn$ and phase space arguments of the functions $O(x,p), Q(x,p)$ on the right hand side for simplicity)
\begin{eqnarray}
[O \circ Q]_{nn} (x,p) &\approx&  O(x,p) Q(x,p)
+ \frac{i}{2} \boldsymbol \Omega \cdot (\nabla_{\v x}  O \times \nabla_{\v x} Q)
\notag \\
&&
-\frac{i}{2} \left (\partial_t O \partial_{E} Q - \partial_{E} O \partial_t Q \right )
+\frac{i}{2} (\nabla_{\v x}  O \nabla_{\v p} Q - \nabla_{\v p} O \nabla_{\v x} Q ). \label{eq:PoissonBerry}
\end{eqnarray}
We use the symbol 
`$\circ$' to denote integration/summation over repeated indices. Apart from the classical Poisson brackets we highlight the appearance of the Berry curvature in the complex, antisymmetric terms.

\subsection{Dyson equation}

\subsubsection{General form and strategy}

Following the standard strategy~\cite{KamenevBook}, the Boltzmann equation is derived systematically from the Keldysh component of the Dyson equation (disorder average is denoted $\langle \dots \rangle$)
\begin{equation}
- \left(\big[G^{R}\big]^{-1} \circ F- F\circ \big[G^{A}\big]^{-1}\right) = \left\langle \Sigma^K - \big(\Sigma^R \circ F - F \circ \Sigma^A\big) \right\rangle. 
\label{eq:DysonMatrix}
\end{equation}
Each of the operators entering Eq.~\eqref{eq:DysonMatrix} is a matrix in $N\times N$ band space, and in space-time (or equivalently, after Fourier transformation, in energy-momentum space). For example, the inverse retarded bare Green's function is $[G^{R}(p,p')]^{-1} = [E + i\eta - H_{\rm kin} (\v p)] (2\pi)^{d + 1} \delta( p - p')$. The operator $F$ is the unknown of the equation and, as usual, is introduced to parametrize the Keldysh component of the Green's function
\begin{equation}
G^K = G^R  \circ F - F \circ G^A.    
\end{equation}
The strategy to find a solution to Eq.~\eqref{eq:DysonMatrix} is a semiclassical expansion using the Wigner-transform $F(x,p)$ of which the on-shell intraband components define the distribution function.

\begin{figure}[t!]
\begin{center}
\includegraphics[width  = .8 \textwidth]{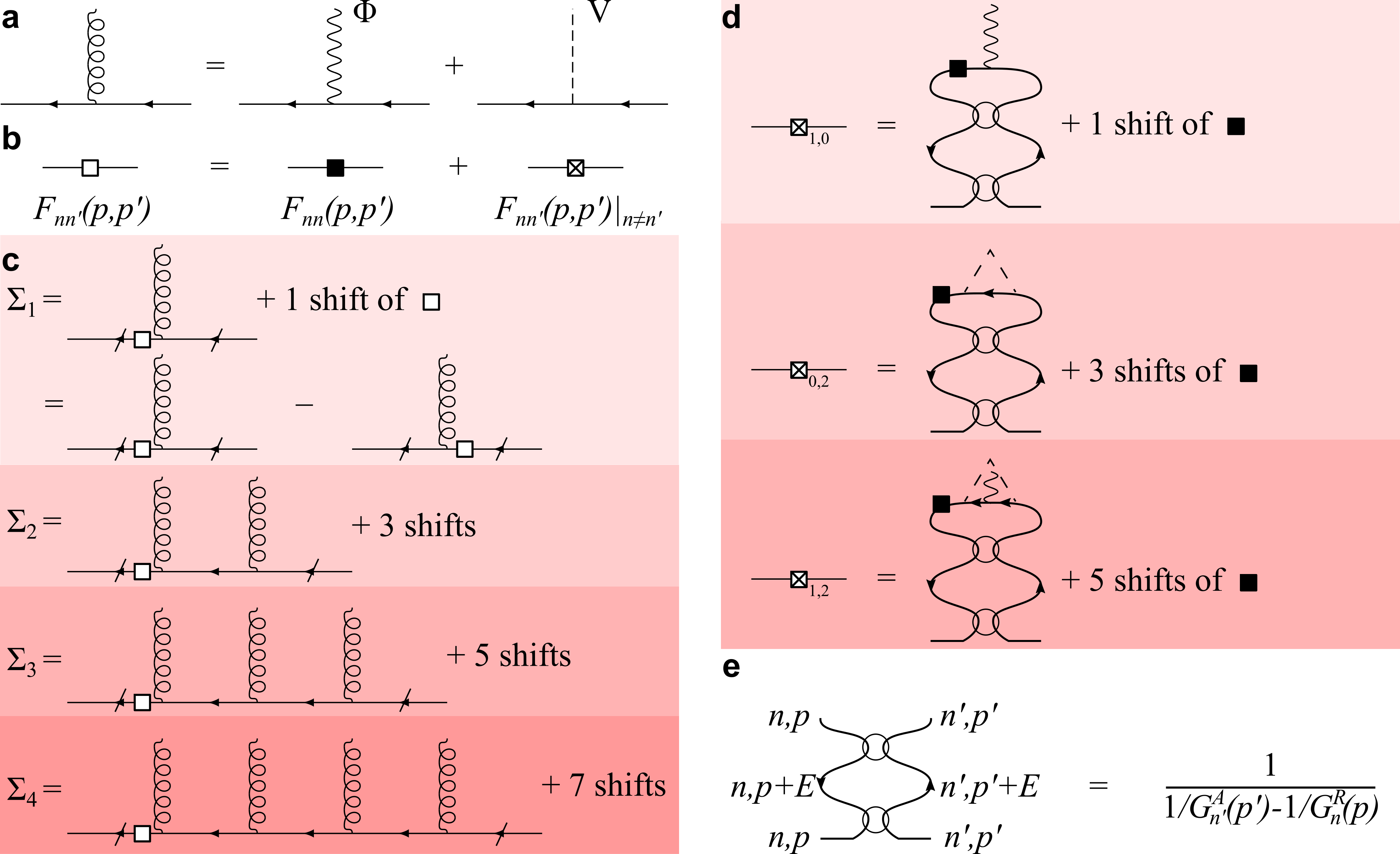}
\caption{Illustration of the diagrammatic rules explained in the text, including the external potential (panel {\bf a}) and the Keldysh operator $F$ (panel {\bf b}). {\bf c} Diagrammatic representation of the self energy up to fourth order in powers of $U = \Phi + V$. {\bf d} Diagrammatic representation of all band off-diagonal contributions to $F_{nn'}(p,p')$. {\bf e} Definition of the vertex leading to the denominator $[G^{A}_{n'}(p')]^{-1}-[G^{R}_n(p)]^{-1}$ (as obtained by taking an energy integral over $E$ and assuming that the upper loop contains one $F$ function, such that the two Green's functions inside the loop have different advanced/retarded structure). Here and in all diagrams, a slashed propagator represents an external on-shell leg.}
\label{fig:DefDiagrams}
\end{center}
\end{figure}

\subsubsection{Self-energy}
\label{sec:SelfEnergy}

The right hand side of Eq.~\eqref{eq:DysonMatrix} contains the self-energy and thereby the effective impact of the disorder potential after averaging. We also incorporate the scattering of the external electric field in the self-energy. A diagrammatic representation, before disorder average $\langle \dots \rangle$, is given in Fig.~\ref{fig:DefDiagrams}, we use the notation $\langle \Sigma^K - (\Sigma^R \circ F - F \circ \Sigma^A) \rangle \simeq \sum_{i = 1}^4 \Sigma_i$ where $i$ counts the order in external perturbation. The following Feynman rules are used.
\begin{enumerate}
\item An arrow on a solid line represents a Green's function $G^{R/A}_n$.
\item An empty square on the solid line represents $F_{nn'}(p,p')$, which contains band diagonal (solid box) and off-diagonal (crossed box) contributions. Per definition, all Green's functions with arrow towards (away from) the square are advanced (retarded). 
\item A curly line represents $U(p,p')= \Phi(p,p') + V(p,p')$, where the photon is represented by a wavy line and disorder by a dashed line.
\item To $m$th order draw all diagrams for the self-energy in the presence of $m$ curly potential lines connected by $m -1$ Green's functions. External legs are on-shell, which is represented by a dashed arrow.
\item Generate $2m$ diagrams by placing the empty square between all vertices and arrows (including those which are dashed). The sign of the diagram is $(-1)^s$ where $s =$(number of vertices downstream)$+$(number of arrowheads downstream), see Fig.~\ref{fig:DefDiagrams} {\bf b} for an example.  
\item Take the disorder average and, as usual, keep only one particle irreducible diagrams and integrate over internal momenta.
\end{enumerate}

\subsubsection{Interband Keldysh function}

While the on-shell part to intraband (i.e.~diagonal) matrix elements of $F(x,p)$ corresponds to the distribution function of states in a given band, off-diagonal matrix elements $F_{nn'}$ ($n \neq n'$) are inherently quantum mechanical and appear perturbatively in powers of $V, \Phi$. We thus derive off-diagonal contributions to $F$ order by order, for a diagrammatic representation see Fig.~\ref{fig:DefDiagrams}. In the self-energy we use the subscript $\Sigma_{[l,m]}$ to denote $l$th order in $\boldsymbol{\mathcal{E}}$ and $m$th order in $V$. We drop the order $l =2$ which contains two small denominators representing virtual transitions.

\begin{figure}[t!]
\begin{center}
\includegraphics[width  = .8 \textwidth]{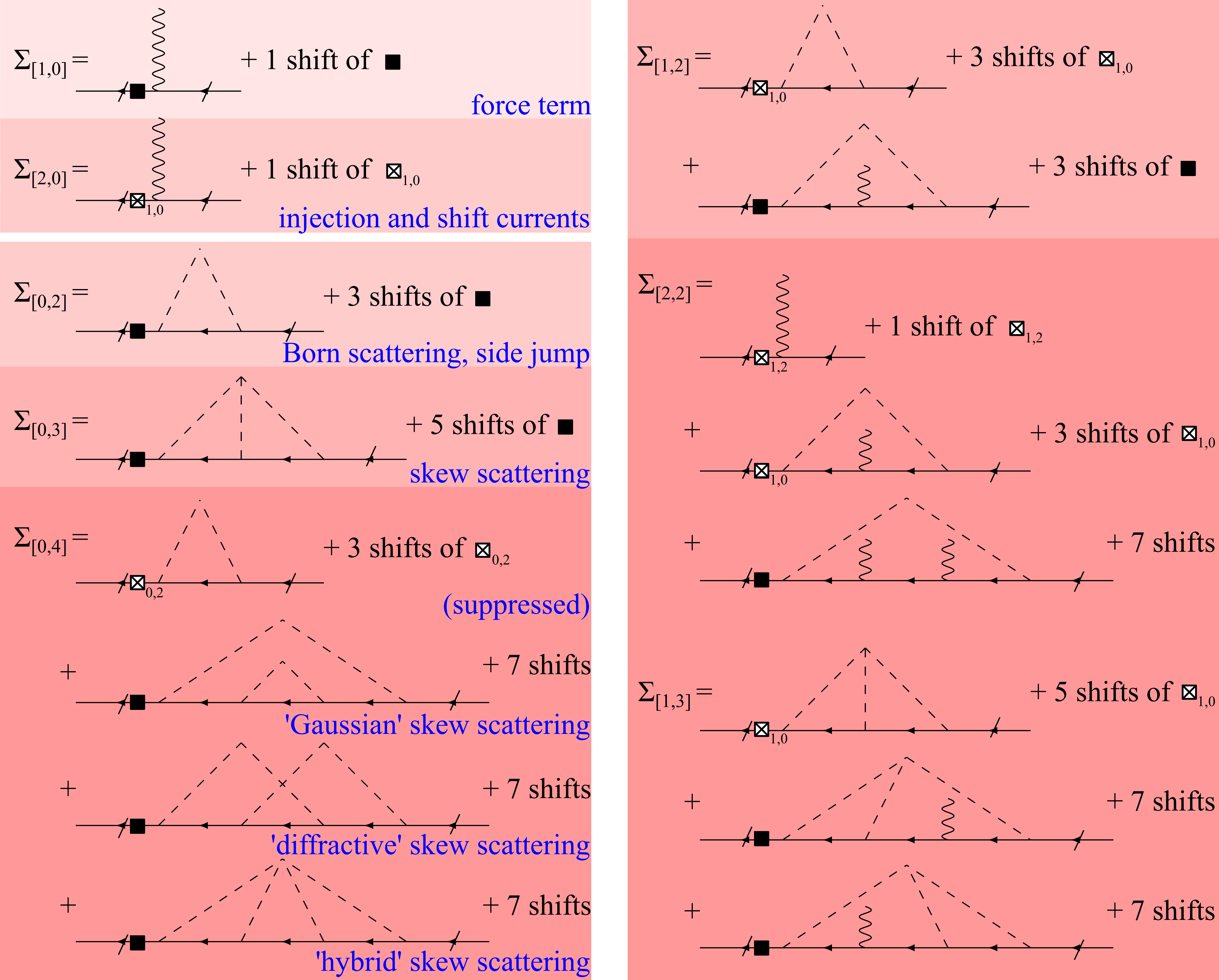}
\caption{Effective total self energy for the intraband projected Dyson equation, Eq.~\eqref{eq:DysonDiago}, keeping only terms which are either due to photons [top left panel], only, or due to disorder, only [bottom left panel]. [Right panel]: Mixed contributions due to the self energy due to both disorder and photons.}
\label{fig:Sigmas}
\end{center}
\end{figure}

\subsubsection{Effective Dyson equation}
We insert this approximate solution for the off-diagonal terms into Eq.~\eqref{eq:DysonMatrix} in order to perturbatively diagonalize the equation. 
The result is 
\begin{equation}
-\frac{1}{2}\left [\big[G_n^{R}\big]^{-1}+\big[G_n^{A}\big]^{-1}\KC F_{nn} \right](p,p') = \Sigma_n(p,p').\label{eq:DysonDiago}
\end{equation}
Here, the symbol $\circ$ denotes space-time (or energy momentum) integration (without band summation) and $[ \dots \KC \dots]$ is the corresponding commutator. We will henceforth omit the band index $n$ whenever possible. The effective self-energy \begin{equation}
\Sigma(p,p') =  \sum_{\substack{l,m = 1\\l+ m \leq 4}}^4 \Sigma_{[l,m]}(p,p')
\end{equation}
can be Taylor expanded, where $\Sigma_{[l,m]}(p,p')$ is $l$th order in $\boldsymbol{\mathcal E}$ and $m$th order in $V$. We keep up to second order in electric field (as necessary for nonlinear transport) and up to fourth order in impurity scattering (which is important for skew scattering). Combined contributions are kept up to $l + m  = 4$. The various contributions to $\Sigma_{l,m}$ are represented in Fig:~\ref{fig:Sigmas}, where intraband components $F_{nn}(p,p')$ are represented by a black box.

\subsection{Semiclassical expansion for $\omega \ll E_F$}

As announced above, the Boltzmann equation follows from Eq.~\eqref{eq:DysonDiago} by a semiclassical gradient expansion. We first focus on the small frequency regime $\omega \ll E_F$. In this regime it is convenient to absorb the external potential into the Green's function $\mathcal{G}^{R/A} = E \pm i\eta - H_{\rm kin}(\v p) - \Phi(x)$. This implies that $\Sigma_{[1,0]}$ is incorporated into the left hand side of Eq.~\eqref{eq:DysonDiago}. We note that interband transitions due to photons are suppressed at small frequencies, such that $\Sigma_{[2,0]}$ is neglected along with all mixed contributions, Fig.~\ref{fig:Sigmas}, with crossed boxes. All other diagrams of Fig.~\ref{fig:Sigmas} (right panel) are accounted for by replacing Green's functions $G^{R/A}$ by $\mathcal{G}^{R/A}$ in Fig.~\ref{fig:Sigmas} (left panels). 
We define the distribution function by enforcing the on-shell constraint on diagonal components $F_{nn}(x,p)$
\begin{equation}
f_{n} (t, \v x, \v p) = \int_E \delta (E - E_n(\v p) - \Phi(x)) F_{n n}(x,p). \label{eq:DistrFunctionSlow}
\end{equation}

\subsubsection{Liouville term}

Using the Moyal expansion derived in Eq.~\eqref{eq:PoissonBerry} we readily obtain the Liouvillian term describing the reactive response of the system
\begin{align}
-\frac{1}{2}\left[\big[\mathcal{G}_n^{R}\big]^{-1}+\big[\mathcal{G}_n^{A}]^{-1} \KC F_{n n} \right](x,p) &\approx
-i \Big ([\partial_t+ \partial_t \Phi \partial_E] F_{n n} (x,p) \notag \\
&+ [\v v_n-\boldsymbol \Omega_n \times\nabla_{\v x} \Phi]\cdot \nabla_{\v x} F_{n n}(x,p)- \nabla_{\v x} \Phi\cdot\nabla_{\v p} F_{n n}(x,p)\Big ),
\end{align}
where the band velocity is $\v v_n =\nabla_{\v p} E_n(\v p)$. Clearly it follows then 
\begin{align}
    -\frac{i}{2}\int_E \delta (E - E_n(\v p) - \Phi(x)) 
    \left[[\mathcal{G}_n^{R}]^{-1}+[\mathcal{G}_n^{A}]^{-1} \KC F_{n n} \right](x,p)
    \approx \left (\partial_t + \dot{\v p} \cdot \nabla_{\v p} + \dot{\v x} \cdot  \nabla_{\v x}\right ) f_n(t, \v x, \v p),\label{eq:LowFreqLiouville}
\end{align}
which reproduces the usual appearance of the Liouville term in the Boltzmann equation, where $\dot{\v x} = \v v_n + \boldsymbol{\Omega}_n \times \dot {\v p}$, and $\dot {\v p} = -\nabla_{\v x} \Phi$.

\subsubsection{Born scattering and side jump}

The diagram for these processes is depicted in Fig.~\ref{fig:Sigmas} (top left panel) and its analytical expression in energy-momentum space is
\begin{align}
\Sigma_{[0,2]}(p_+,p_-) &= n_{\rm imp} \oint_{n'\v p'} F_{nn}(p_+, p_-) \vert \mathcal{V}((\v p - \v b_n) - (\v p' - \v b_{n'}))\vert^2  \notag \\ 
&\times\left[\vert \braket{u_{n, \v p_-} \vert E \vert u_{n', \v p'_-}} \vert^2\mathcal{G}_{n'}^A(E_-, \v p_- ')- \vert \braket{u_{n, \v p_+} \vert E \vert u_{n', \v p'_+}} \vert^2 \mathcal{G}_{n'}^R(E_+, \v p_+')
\right] \notag \\ 
&-n_{\rm imp}\oint_{n'\v p'} F_{n'n'}(E_+, \v p_+'; E_-, \v p_-')\vert \mathcal{V}((\v p - \v b_n) - (\v p' - \v b_{n'})) \notag \\ 
&\times\braket{u_{n, \v p_+} \vert E \vert u_{n', \v p'_+}}\braket{u_{n', \v p_-'} \vert E \vert u_{n, \v p_+}}\vert^2 [ \mathcal{G}_{n'}^A(E_-, \v p_- ') - \mathcal{G}_{n'}^R(E_+, \v p_+ ')].
\end{align}
Here we introduced a joint integration symbol $\oint$ that implies summation over the discrete index $\sum_{n'}$ and integration over the continuous variable $\int_{\v p'}$.  We expand the above expression in small $\Delta \v p$ and use the Wigner transform, to obtain
\begin{align}
\Sigma_{[0,2]}(x,p) &\approx i n_{\rm imp} \oint_{n'\v p'} (2\pi)^2 \delta (E - E_{n'}(\v p')) \delta(E - E') \vert \mathcal{V}((\v p - \v b_n) - (\v p' - \v b_{n'}))\vert^2 \vert \braket{u_{n, \v p} \vert E \vert u_{n', \v p'} } \vert^2 \notag \\
&\times\Big [ F_{nn}(x, p) - [1 + \delta \v r_{n'n}(\v p', \v p) \nabla_{\v x}] F_{ n' n'}(x,p')\Big ]. \label{eq:Sigma02}
\end{align}
Here, we dropped principle value integrals (suppressed virtual processes) and introduced the displacement at a side-jump ~\cite{BelinicherSturman1982,SinitsynMacDonald2006}
\begin{align}
\delta \v r_{n' n}(\v p', \v p) = \boldsymbol{\mathcal A}_{n'} (\v p')- \boldsymbol{\mathcal A}_n(\v p)- (\nabla_{\v p} + \nabla_{\v{p}'}) \arg (\braket{u_{n,' \v p'} \vert E \vert u_{n, \v p}}).
\end{align}
After projecting onto the mass shell, we obtain the collision integral (``Stossintegral'' in German) of the right-hand-side in the Boltzmann equation 
\begin{align}
\mathrm{St}_{\rm Born}\{f\} + \mathrm{St}_{\rm sj}\{f\} &= -2\pi  n_{\rm imp}\oint_{n'\v p'}
\delta (E_n(\v p) +\Phi(t, \v x) - E_{n'}(\v p') - \Phi(t, \v x + \delta \v r_{n' n}(\v p' ,{\v p}))) \notag \\
&\vert \mathcal{V}((\v p - \v b_{n}) - (\v p' - \v b_{ n'}))\vert^2 \vert \braket{u_{n, \v p} \vert E \vert u_{n', \v p'}} \vert^2 
\left [f_n(t, \v x, \v p)-f_{n'}(t, \v x + \delta \v r_{n'n}(\v p' ,{\v p}),\v p')\right ].\label{eq:LowFreqI02}
\end{align}
We highlight that this expression, which contains both Born scattering $\mathrm{St}_{\rm Born}\{f\}$ (Eq.~\eqref{eq:LowFreqI02} at $\delta \v r = 0$) and the side-jump effect $\mathrm{St}_{\rm sj}\{f\}$ (Eq.~\eqref{eq:LowFreqI02} to first order in $\delta \v r = 0$), is to be understood up to first order in the gradient expansion. We remind that in common notations, the latter collision term $\mathrm{St}_{\text{sj}}\{f\}$ is often referred to as anomalous distribution.   

\subsubsection{Skew scattering}
It will be useful to use a multi-index notation $l = (n, \v p)$ as well as energy notation $E_l = E_n(\v p)$, and introduce matrix elements $V_{ll'} = \braket{u_{n, \v p} \vert E \vert u_{n', \v p'}} {V}((\v p - \v b_n) - (\v p' - \v b_{n'}))$. In this notation, we find prior to disorder average
\begin{align}
    \Sigma_{[0,3]}(x,p) &=\oint_{l_1, l_2} F_{nn}(x,p)  V_{ll_1}V_{l_1 l_2} V_{l_2l } \left[\mathcal{G}_{l_1}^A\mathcal{G}_{l_2}^A - \mathcal{G}_{l_1}^R\mathcal{G}_{l_2}^R\right] \notag \\
&+F_{n_1n_1}(x, p_1) \left[\mathcal{G}_{l_1}^R-\mathcal{G}_{l_1}^A\right]
\left[V_{ll_1}V_{l_1 l_2} V_{l_2l } \mathcal{G}_{l_2}^A +V_{ll_2}V_{l_2 l_1} V_{l_1l}\mathcal{G}_{l_2}^R \right].
\end{align}
We emphasize that all Green's function have the same energy argument since impurity scattering is elastic. We only need contributions which are odd under $l \leftrightarrow l_1$ and obtain, after on-shell projection, the skew scattering collision term in the form 
\begin{subequations}
\begin{equation}
\mathrm{St}_{\text{sk}}\{f\} = \oint_{l'} W_{ll'}^{\text{sk}} f_{l'},
\end{equation}
where the corresponding transition probability is 
\begin{align}
     W_{ll'}^{\text{sk}} = -\oint_{l_2}(2\pi)^2 \delta (E_l - E_{l'}) \delta(E_{l} - E_{l_2})  \mathrm{Im}\big[V_{ll'}V_{l'l_2} V_{l_2l}\big] .  
\end{align}\label{eq:LowFreqI03}

With additional simplifying assumptions further analytical progress can be made. For instance, one tractable example corresponds to a disorder model created by a centrosymmetric impurity potential, $\mathcal V(\v x)=\mathcal V(-\v x)$, that is assumed to vary slowly on the scale of the lattice constant. For this model, the electron transition matrix element separates into the product of a Fourier transform of the impurity potential and a Bloch wave function overlap of states within the same node, $\propto \mathcal V({\v p_1-\v p_2})\braket{u_{l_1} \vert u_{l_2}} \simeq V_{0}\braket{u_{l_1} \vert u_{l_2}}$, where $V_{0}$ denotes the strength of intranode scattering. Since for a centrosymmetric impurity $\mathrm{Im}[V_{\v p}]=0$, the antisymmetric part of the scattering probability defined by Eq.~\eqref{eq:LowFreqI03} becomes after disorder average 
\begin{align}
  W_{ll'}^{\text{sk}} = -n_{\rm imp} V_0^3\oint_{l''}(2\pi)^2 \delta (E_l - E_{l'}) \delta(E_{l} - E_{l''})  Z_{ll'l''},
\end{align}
where
\begin{equation}\label{eq:Z}
    Z_{l_1l_2l_3} = \text{Im} [\braket{u_{l_1} \vert u_{l_2}}\braket{u_{l_2} \vert u_{l_3}}\braket{u_{l_3} \vert u_{l_1}}],
\end{equation}
This contribution appears from third order scattering of a single impurity, see Fig.~\ref{fig:Sigmas}.
\end{subequations}

\subsubsection{Gaussian, diffractive, and hybrid skew scattering}

The first line of $\Sigma_{[0,4]}$ in Fig.~\ref{fig:Sigmas} is small in powers of the semiclassical parameter of Fermi energy being much larger than the elastic scattering rate. This follows, because both the crossed box and the Green's function account for virtual elastic (i.e. horizontal) interband transitions. Prior to disorder average, we thus find
\begin{align}
\Sigma_{[0,4]}(x,p) &= \oint_{ l', l_1, l_2} F_{nn}(x,p) V_{l l'}V_{ l' l_1}V_{l_1 l_2}V_{l_2l}[\mathcal G_{ l'}^A\mathcal G_{l_1}^A\mathcal G_{l_2}^A- \mathcal G_{ l'}^R\mathcal G_{l_1}^R\mathcal G_{l_2}^R]  +F_{ n' n'}(x, p') [\mathcal G^R_{ l'} - \mathcal G^A_{ l'}] \notag \\ 
 &
 \times\lbrace  V_{l l'}V_{ l' l_1}V_{l_1 l_2}V_{l_2l}\mathcal G_{l_1}^A \mathcal G_{l_2}^A +V_{l l_1}V_{l_1 l_2}V_{l_2 l'} V_{ l' l}\mathcal G_{l_1}^R \mathcal G_{l_2}^R + V_{l l_1}V_{l_1  l'}V_{ l' l_2} V_{l_2 l}\mathcal G_{l_1}^R \mathcal G_{l_2}^A\rbrace.
\end{align}
Again, we only keep the contribution which is odd under $l \leftrightarrow  l'$ and project on-shell to obtain
\begin{subequations}
\begin{equation}
\mathrm{St}_{\rm sk}^{[4]}[f] =\oint_{ l'} \widetilde{W}_{l l'}^{\rm sk} f_{ l'},
\end{equation}
where
\begin{align}
    \widetilde{W}_{l l'}^{\rm sk} &= - \oint_{l_1,l_2}(2\pi)^2 \delta (E_l - E_{l_1}) \delta(E_{l} - E_{ l'}) [\mathcal G_{l_2}^R + \mathcal G_{l_2}^A]/2 \notag \\
     & \times\left\lbrace  \mathrm{Im}[V_{l l'}V_{ l' l_1}V_{l_1 l_2}V_{l_2l} + (1 \leftrightarrow 2)] + 2 \mathrm{Im}[V_{l l_2}V_{l_2  l'} V_{ l' l_1} V_{l_1 l}] \right\rbrace.  
\end{align}\label{eq:LowFreqI04}
Note that contrary contrary to all previous contributions, here one off-shell contribution $[\mathcal G_{l_2}^R + \mathcal G_{l_2}^A]$ is explicit. Disorder average implies the impurities leading to three different kinds of diagrams, see Fig.~\ref{fig:Sigmas}: scattering from two different impurities allows for a rainbow diagram (Gaussian skew scattering)~\cite{SinitsynMacDonald2006} and a crossed diagram (diffractive skew scattering)~\cite{AdoTitov2015,KoenigLevchenko2016}, while scattering from a single impurity to fourth order leads to hybrid skew scattering~\cite{KovalevSinova2008}. 

We illustrate these three contributions, $\widetilde{W}_{l l'}^{\rm sk} = W_{l l'}^{\rm Gauss} + W_{l l'}^{\rm diff} + W_{l l'}^{\rm hybrid}$ again in the limit of a smooth, centrosymmetric impurity potential of intranode scattering strength $V_0$. After impurity average, the expressions simplify to
\begin{align}
     W_{l l'}^{\rm Gauss} &= - (n_{\rm imp} V_0)^2 \oint_{l_1, l_2}2\pi^2 \delta (E_l - E_{l_1}) \delta(E_{l} - E_{ l'})  [\mathcal G_{l_2}^R + \mathcal G_{l_2}^A] \notag \\ &\times\lbrace( Z_{ll'l_1 l_2} \delta_{\v p_2, \v p'}+ Z_{ll'l_2 l_1} \delta_{\v p_1, \v p'} )+ 2 Z_{ll_2l' l_1} \delta_{\v p_1, \v p_2} \rbrace,  \\
     W_{l l'}^{\rm diff} &= - (n_{\rm imp} V_0)^2\oint_{l_1, l_2}2\pi^2 \delta (E_l - E_{l_1}) \delta(E_{l} - E_{ l'})
     [\mathcal G_{l_2}^R + \mathcal G_{l_2}^A] \notag \\ &\times \lbrace( Z_{ll'l_1 l_2}\delta_{\v p_2- \v p_1, \v p - \v p'}  + 1 \leftrightarrow 2) )+ 2 Z_{ll_2l' l_1} \delta_{\v p_1+ \v p_2, \v p + \v p'}\rbrace,  \\
      W_{l l'}^{\rm hybrid} &= - n_{\rm imp} (V_0)^4\oint_{l_1, l_2}2\pi^2 \delta (E_l - E_{l_1}) \delta(E_{l} - E_{ l'})[\mathcal G_{l_2}^R + \mathcal G_{l_2}^A] \lbrace( Z_{ll'l_1 l_2} + 1 \leftrightarrow 2 )+ 2 Z_{ll_2l' l_1} \rbrace,  
\end{align}
where we introduce the four index analog to Eq.~\eqref{eq:Z}
\begin{equation}\label{eq:Z4}
    Z_{l_1,l_2,l_3,l_4} = \text{Im} [\braket{u_{l_1} \vert u_{l_2}}\braket{u_{l_2} \vert u_{l_3}}\braket{u_{l_3} \vert u_{l_4}}\braket{u_{l_4} \vert u_{l_1}}].
\end{equation}
We conclude the section on skew scattering from the fourth order potential with a technical remark relating to the diagrammatic calculation of the anomalous Hall conductivity bubble. In the equation for $W_{l l'}^{\rm diff}$, the first two terms in the curly brackets represent so called $\Psi$ diagrams, while the last term is the $X$ diagram (in the notation of \cite{AdoTitov2015,KoenigLevchenko2016,Milletari-Ferreira-PRB16,Ado-PRL16}).
\end{subequations}

\subsubsection{Summary small frequency result}
This concludes the derivation of the Boltzmann equation in Berry curved matter in the low-frequency ($\omega \ll E_F$ limit):
\begin{equation}
    \left[\partial_t  + \dot{\v x} \cdot \nabla_{\v x} + \dot{\v p} \cdot  \nabla_{\v p}\right]f(t,\v x, \v p) = \mathrm{St}\{f\}, \label{eq:BoltzmannSmallFreq}
\end{equation}
where $\dot{\v x} = \v v_n + \boldsymbol{\Omega}_n \times \dot {\v p}$, $\dot {\v p} = - \nabla_{\v x} \Phi$ and the collision integral $\mathrm{St}[f] =\mathrm{St}_{\rm Born}\{f\}  + \mathrm{St}_{\rm sj}\{f\} + \mathrm{St}_{\rm sk}\{f\}  + \mathrm{St}_{\rm sk}^{[4]}\{f\}$ is given by Eqs.~\eqref{eq:LowFreqI02}, \eqref{eq:LowFreqI03}, \eqref{eq:LowFreqI04}. We explicitly kept non-linear orders of the external potential in the collision integral (which eventually vanish), but restricted ourselves to terms of zeroth and first order in gradients $(\hbar \partial_{\v x}\partial_{\v p})$, and dropped the combination of skew and side-jump effects. 

\subsection{Semiclassical expansion for $\omega \gtrsim E_F$}

In this section we derive the effective kinetic theory for an external field with fast driving frequency. We concentrate on the rectified current stemming from slowly fluctuating $F_{nn}(t, \v x; E, \v p)$ for which a semiclassical gradient expansion is justified, while we omit quickly oscillating first and second harmonics of $F_{nn}(t, \v x; E, \v p)$. The physical reason behind concentrating on the rectified current within the present Boltzmann technique is that the latter is mainly designed to incorporate relaxation effects. On the other hand, relaxation is negligible for the quickly oscillating first and second harmonic.

Technically, we time average Eq.~\eqref{eq:DysonDiago} $\langle \dots \rangle_{\rm time} = \int_0^{2\pi/\omega} [dt ...]\omega/2\pi$, whereby the terms $\Sigma_{1,0},\Sigma_{1,2},\Sigma_{1,3}$ vanish. Since the external potential is quickly oscillating, we use a slightly different definition of the distribution function than in the slow frequency limit (cf. Eq.~\eqref{eq:DistrFunctionSlow})
\begin{equation}
f_{n} (t, \v x, \v p) = \int_{E} \delta (E - E_n(\v p)) F_{n n}(x,p). \label{eq:DistrFunctionFast}
\end{equation}

\subsubsection{Liouville term and disorder scattering}

The Liouville term and disorder induced collision integrals for the rectified distribution function are essentially the same as in the slow-frequency limit, Eqs.~\eqref{eq:LowFreqLiouville}, \eqref{eq:LowFreqI02}, \eqref{eq:LowFreqI03}, \eqref{eq:LowFreqI04}, and follow from analogous derivations. The only major difference is the cancellation of linear terms in the external potential leading to absent force term, $\dot {\v p} = 0$, and side jump contributions.

\subsubsection{Injection and shift currents}

The vertical interband scattering off two photons leads to a self-energy contribution
\begin{align}
\Sigma_{[2,0]}(x,p) \simeq  i\sum_{n'} \sum_{i,j = 1}^d\sum_{\xi = \pm} 2\pi  \delta (\xi \omega + E_{n'}(\v p) - E_{n}(\v p)) 
e^2 \mathcal E_\xi^i \mathcal E_{-\xi}^j \mathcal A_{nn'}^i(\v p)\mathcal A_{n'n}^j(\v p) \notag \\ \times \Big \lbrace F_{nn}(x, p) 
- [1 + \v R_{n'n}(\v p)\cdot \nabla_{\v x}] F_{n'n'}(x,p - \xi \omega)\Big \rbrace.
\end{align}
Here we used a suggestive notation $p - \xi \omega = (E - \xi \omega, \v p)$. On a technical level, the derivation is very similar to the derivation of Eq.~\eqref{eq:Sigma02} presented above. Again, the scattering process is accompanied by a coordinate shift, which for photons reads~\cite{BelinicherSturman1982}
\begin{align}
\v R_{n' n}(\v p) = \boldsymbol{\mathcal A}_{n'} (\v p)- \boldsymbol{\mathcal A}_n(\v p) - \nabla_{\v p} \arg(\boldsymbol{\mathcal A}_{n' n}(\v p)). 
\end{align}
The projection of this equation on the mass shell generates a collision ``integral'' (a source term for photocarrier injection) 
\begin{align}
 \mathrm{St}_{\rm PGE}\{f\} = -\sum_{n'} \sum_{i,j = 1}^d\sum_{\xi = \pm} 2\pi  \delta (\xi \omega +E_{n'}(\v p) - E_{n}(\v p))
 e^2 \mathcal E_\xi^i \mathcal E_{-\xi}^j \mathcal A_{nn'}^i(\v p)\mathcal A_{n'n}^j(\v p) \notag \\ 
 \times\Big \lbrace f_n(t, \v x,  \v p)- [1 + \v R_{n'n}(\v p)\cdot \nabla_{\v x}] f_{n'}(t, \v x, \v p)\Big \rbrace. \label{eq:LargeFreqInjection}
\end{align}

\subsubsection{Mixed disorder and photon scattering}

The interplay of disorder and optical excitation is encoded in $\Sigma_{[2,2]}(x,p)$. To leading order, there are two impurity induced contributions for optical transitions.

First, a particle in the filled valence band first scatters off an impurity to a virtual off-shell state, and then performs an optically allowed transition. Second, the reverse happens, i.e. an optical transition to an off-shell state in the conduction band, and an impurity assisted scattering to an on-shell final state in the conduction band. The contribution $\Sigma_{[2,2]}(x,p)$ describes these transitions and we leave a careful study to a separate publication.

We remark that $\Sigma_{[1,3]}(x,p)$ vanishes upon time average (as mentioned, this is the only relaxation prone channel in the large frequency limit). Therefore interband contributions linear in external field, as discussed in Ref.~\cite{XiaoNiu2019}, are disregarded.

\subsubsection{Summary of large frequency result}

In total, the kinetic equation at large frequencies takes the form
\begin{equation}
    [\partial_t  + \dot{\v x} \cdot \nabla_{\v x}]f(t,\v x, \v p) = \mathrm{St}\lbrace f\rbrace, \label{eq:BoltzmannLargeFreq}
\end{equation}
where $\dot{\v x}= \v v_n$ and $\mathrm{St}\lbrace f\rbrace$ has contributions from the injection of photocarriers, Eq.~\eqref{eq:LargeFreqInjection}, relaxation and skew scattering effects at impurities, Eqs.~\eqref{eq:LowFreqI02}, \eqref{eq:LowFreqI03}, \eqref{eq:LowFreqI04}, as well as mixed contributions to be discussed elsewhere.

Note that the kinetic Eq.~\eqref{eq:BoltzmannLargeFreq} contains a carrier injection far away from the Fermi surface - here the energy of on-shell particles does not necessarily coincide with the Fermi level and interactions induce inelastic energy relaxation effects necessary for photocarrier decay.


\begin{figure}[t!]
\begin{center}
\includegraphics[width=.8\textwidth]{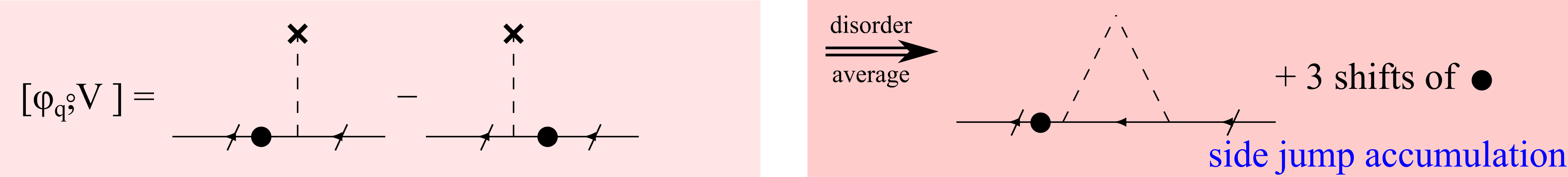}
\caption{Renormalization of the current vertex by the impurity potential. In the disorder contribution to the commutator Eq.~\eqref{eq:deltaGPhiq}, the U(1) phase $\phi_{\text{q}}$ is represented by a disk.}
\label{fig:SideJumpAccumulation}
\end{center}
\end{figure}

\section{Electrical current response}\label{sec:Current}

In this section we present derivation of the electrical current response and elucidate the origin of the side-jump accumulation velocity.
 
In the field integral representation of the Keldysh technique~\cite{KamenevBook}, the electrical current is defined by a functional derivative of the partition function, $j_i(x) = [\partial Z/\partial A^i_{\rm q}(x) ] \vert_{\v A_{\text q} = 0}$, where the index-$\text{q}$ indicates the quantum component of the vector potential. The coupling of the latter to the fields is imposed by electromagnetic U(1) gauge invariance. Therefore, the current operator associated to Bloch-electrons of band $n$ can be obtained from the gauge transformation of 
\begin{equation}
\hat G^{-1}_{nn} = \left (\begin{array}{cc}
[\hat{G}^{R}]^{-1} & [\hat G^{-1}]^{K} \\ 
0 & [\hat{G}^{A}]^{-1}
\end{array}  \right)_{nn} \rightarrow e^{- i \phi_{\text{q}}(x) \hat \gamma_{\text{q}}}\hat G^{-1}_{nn} e^{i \phi_{\text{q}}(x) \hat \gamma_{\text{q}}}.
\end{equation}
Here, following standard convention, $\hat \gamma_{\text{q}}$ is the first Pauli matrix in Keldysh space. To leading order
\begin{equation}
\delta \hat G^{-1}_{nn} = i \left [ \phi_{\text{q}} \KC \hat H_{nn}\right ] \hat \gamma_{\text{q}}, \label{eq:deltaGPhiq}
\end{equation}
where $\hat H_{nn}$ is the full Hamiltonian prior to disorder average and projected on band $n$. We omitted terms which are stemming from the regulation of Keldysh theory (e.g. $[\hat G^{R}]^{-1} - [\hat G^{A}]^{-1} \sim i \eta \rightarrow 0$).

For the clean part of $\hat H_{nn}$ we use the Wigner transform Eq.~\eqref{eq:GeneralizedWigner} to obtain $\delta \hat G^{-1}_{nn}(x,p) \vert_{V = 0} \approx - \dot{\v x} \cdot \nabla_{\v x}{\phi_q} \hat \gamma_{\text{q}}$, where $\dot{\v x}  = \v v  + \boldsymbol \Omega \times \dot{\v p}$. The contribution from the disorder potential contains a renormalization of the vertex depicted diagrammatically in Fig.~\ref{fig:SideJumpAccumulation}. Here, the matrix $\hat \gamma_{\text{q}}$ imposes that Green's functions which are coming into (going out of) the vertex are advanced (retarded). The similarity of this diagrams with $\Sigma_{[0,2]}$ in  Fig. \ref{fig:Sigmas}, where the square representing $F_{nn}(x,p)$ is replaced by the disk representing $\phi_q(x)$, implies that the vertex renormalization is given in complete analogy to Eq.~\eqref{eq:Sigma02} by 
\begin{align}
\langle\delta \hat G^{-1}_{nn}(x,p) \rangle &\approx - n_{\rm imp} \oint_{n'\v p'} (2\pi) \delta (E - E_{n'}(\v p')) \delta \v r_{n'n}(\v p', \v p) \nabla_{\v x} \phi_{\text{q}}\notag \\
& \times\vert \mathcal{V}((\v p - \v b_n) - (\v p' - \v b_{n'}))\vert^2 \vert \braket{u_{n, \v p} \vert E \vert u_{n', \v p'} } \vert^2 \hat \gamma^q. \label{eq:Sidejumpaccumulation}
\end{align}
The average current is thus
\begin{align}
	\v j= {e}
	\int_{\v p} [\dot{\v x} + {\v v}^{\rm sj}] f_{n}(t, \v x, \v p).
\end{align}
Here, we introduced the side jump accumulation velocity 
\begin{align}
\v v^{\rm sj} = n_{\rm imp} \oint_{n'\v p'} (2\pi) \delta (E - E_{n'}(\v p'))  \delta \v r_{n'n}(\v p',\v p)\vert \mathcal{V}((\v p - \v b_n) - (\v p' - \v b_{n'}))\vert^2 \vert \braket{u_{n, \v p} \vert E \vert u_{n', \v p'} } \vert^2. \label{eq:ElectricalCurrent}
\end{align}
In the case of large frequencies, we omit the photon field in $\dot{x}_i$, as it is only important for intraband physics. 


\section{Anomalous Hall transport of multifold fermions}\label{sec:Multifold}

In this section we present an application of our theory for the anomalous transport in the model of multifold fermions, as defined in Eq.~\eqref{eq:Multifold}. 
We remind that for a spin-$S$ the corresponding operators are square matrices of dimension $2S+1$. They can be conveniently represented in the Zeeman basis with states labeling $|S,m\rangle$. We choose to work in this basis.  

\subsection{Eigenstates and Berry curvature}

As the starting point in obtaining the Berry curvature for this model it is convenient to use spherical coordinates for the $\v d$-vector  
\begin{equation} 
\v d = d\, (\sin(\theta)\cos(\phi), \sin(\theta)\sin(\phi), \cos(\theta))^T.
\end{equation}
The eigenstates of Eq.~\eqref{eq:Multifold} then readily follow from rotating the Hamiltonian onto the $\hat z$ axis, i.e.
\begin{equation}
    \ket{u_{m,\v p}} = e^{- i \phi S_z} e^{-i \theta S_y} \hat e_m,
\end{equation}
where $\hat e_m$ is the unit vector pointing in $m$ direction, with the conventional labeling of $m = -S, -S + 1, \dots, S-1, S$. The energy of these states is $E_m(\v p) = d_0 (\v p) + m \,d(\v p)$. From this definition, the Berry curvature can be now computed from the standard formula
\begin{align}
    \Omega_a= i\varepsilon_{abc} \braket{\partial_b u_{m, \v p} \vert \partial_c u_{m, \v p}}
    = m \varepsilon_{abc} \partial_b \phi \partial_c \theta \sin(\theta) = - \frac{m}{2} \varepsilon_{abc} \hat d \cdot (\partial_b \hat d \times \partial_c \hat d).
\end{align}
where the unit vector is $\hat d = \v d/d$.

\subsection{Projectors on eigenstates}
As we shall shortly see, the projectors of eigenstates 
\begin{equation}
\hat{P}_{m,\v p}\equiv\ket{u_{m,\v p}}\bra{u_{m,\v p}} = e^{- i \phi S_z} e^{-i \theta S_y} \hat e_m \hat e_m^T e^{i \theta S_y} e^{i \phi S_z} 
\end{equation}
determine the microscopic form of the scattering rates. Specifically, we here review the simplest cases of $S \leq 3/2$. For the $S=1/2$ case the result is well known 
\begin{equation}\label{eq:Projector-S1}    
\hat{P}_{m,\v p} = \frac{1 + 4 m (\hat d \cdot \v S)}{2}. 
\end{equation}
In contrast, for the $S=1$ one finds instead a different expression 
\begin{equation}\label{eq:Projector-S2}
\hat{P}_{m,\v p} = (1 - m^2) + \frac{m}{2}(\hat d \cdot \v S) + \frac{3 m^2 - 2}{2} (\hat d \cdot \v S)^2. 
\end{equation}
Finally, for the $S=3/2$ situation one needs to distinguish between projections with $|m|=3/2$
\begin{equation}\label{eq:Projector-S3}
\hat{P}_{m,\v p} =- \frac{1}{16} - \frac{m}{36}(\hat d \cdot \v S) +\frac{1}{4} (\hat d \cdot \v S)^2 + \frac{m}{9} (\hat d \cdot \v S)^3,    
\end{equation}
and alternatively with $|m|=1/2$, for which 
\begin{equation}\label{eq:Projector-S4}
\hat{P}_{m,\v p} =\frac{9}{16} + \frac{9 m}{4} \hat d \cdot \v S- \frac{1}{4} (\hat d \cdot \v S)^2-m (\hat d \cdot \v S)^3.     
\end{equation}

\subsection{Overlap of states}

The overlap of states in the same band can be directly calculated in complete analogy to projectors. For the above specified model examples they look as follows. For $S=1/2$
\begin{equation}\label{eq:overlap-1}
  \vert \braket{u_{m,\v p_1}\vert u_{m,\v p_2}}\vert^2 = \frac{1 + \hat d_1 \cdot \hat d_2}{2}.
\end{equation}
For the $S=1$ case the overlap depends on the value of $m$. The simplest is that of $m=0$
\begin{equation}\label{eq:overlap-2}
\vert \braket{u_{m,\v p_1}\vert u_{m,\v p_2}}\vert^2 =  (\hat d_1 \cdot \hat d_2)^2, 
\end{equation}
whereas $|m|=1$ is given by 
\begin{equation}\label{eq:overlap-3}
\vert \braket{u_{m,\v p_1}\vert u_{m,\v p_2}} \vert^2=\frac{(1 + \hat d_1 \cdot \hat d_2)^2}{4}. 
\end{equation}
The overlap for $S=3/2$ with $|m|=1/2$ is more involved
\begin{equation}\label{eq:overlap-4}
\vert \braket{u_{m,\v p_1}\vert u_{m,\v p_2}} \vert^2 = \frac{(1 + \hat d_1 \cdot \hat d_2)^3}{8}  - (\hat d_1 \cdot \hat d_2) (\hat d_1 \times \hat d_2)^2,
\end{equation}
while the last one for $|m|=3/2$ is  
\begin{equation}\label{eq:overlap-5}
\vert \braket{u_{m,\v p_1}\vert u_{m,\v p_2}}\vert^2 =\frac{(1 + \hat d_1 \cdot \hat d_2)^3}{8}.
\end{equation}
In all the above expressions we used a shorthand notation 
$\hat d_i= \v d(\v p_i)/d(\v p_i)$.

\subsection{Pancharatnam phase}

The Pancharatnam phase $\Phi_{\v p_{1}\v p_{2}\v p_{3}}$ of the underlying electronic band structure is defined by the Bloch state overlap of the form 
\begin{equation}\label{eq:PP}
\Phi_{\v p_{1}\v p_{2}\v p_{3}}=\arg[\braket{u_{\v p_1} \vert u_{\v p_2}}\braket{u_{\v p_2} \vert u_{\v p_3}}\braket{u_{\v p_3} \vert u_{\v p_1}}]. 
\end{equation} 
Interestingly, as alluded in the introduction, there is a limit when both the skew scattering rate and the side jump contribution can be directly related to $\Phi_{\v p_{1}\v p_{2}\v p_{3}}$. Indeed, as it follows from the collision terms derived in the preceding sections, the antisymmetric part of the scattering rate depends on both properties of band structure and local impurity potential. Thus generic calculations are possible only based on first principle numerical methods. However, under the simplifying assumptions of a single relevant band and a smooth centrosymmetric potential, we demonstrated that skew scattering depends on the quantitiy
$Z_{\v p_1\v p_2\v p_3}$, Eq.~\eqref{eq:Z} (we use $l_i = (n, \v p_i)$, $i = 1, 2,3$ and suppress the band index $n$) which is non-zero only if the Pancharatnam phase in Eq. \eqref{eq:PP} is finite. 
Below we report results for $Z_{\v p_1\v p_2\v p_3}$ where we retained only intraband contributions. For $S=1/2$ the result was derived earlier~\cite{KoenigPesin2019}
\begin{subequations}
\begin{equation}\label{eq:Z-1}
Z_{\v p_1\v p_2\v p_3} = \frac{m}{2} \hat d_1\cdot (\hat d_2 \times \hat d_3).
\end{equation}
For $S=1$ we find 
\begin{equation}\label{eq:Z-2}
Z_{\v p_1 \v p_2 \v p_2}= \frac{m}{8} \hat d_1 \cdot (\hat d_2 \times \hat d_3)(1 + \hat d_1 \cdot \hat d_2 + \hat d_2 \cdot \hat d_3 + \hat d_3 \cdot \hat d_1).    
\end{equation}
For $S=3/2$ the expressions we get are rather cumbersome 
\begin{align}\label{eq:Z-3}
 Z_{\v p_1\v p_2\v p_3} &= \pm \frac{\hat d_1 \cdot (\hat d_2 \times \hat d_3)}{128} \Big [13 + 12 (\hat d_1 \cdot \hat d_2 + \circlearrowright) - [(\hat d_2 \cdot \hat d_3)^2 + \circlearrowright] \notag \\ 
&+8 (\hat d_1 \cdot \hat d_3 \hat d_2 \cdot d_3 + \circlearrowright) +14 (\hat d_1 \cdot \hat d_2 \hat d_2 \cdot \hat d_3 \hat d_3 \cdot \hat d_1) -9 (\hat d_1 \cdot (\hat d_2 \times \hat d_3))^2\Big]
\end{align}
for $|m|=3/2$ with symbol $\circlearrowright$ indicating cyclic permutation of indices $1 \rightarrow 2 \rightarrow 3 \rightarrow 1$ in the sum of respective terms. Lastly, when $|m|=1/2$ the result is 
\begin{align}\label{eq:Z-4}
  Z_{\v p_1\v p_2\v p_3} &=    
  \pm \frac{\hat d_1 \cdot (\hat d_2 \times \hat d_3)}{128} \Big [-247 + 12 (\hat d_1 \cdot \hat d_2 + \circlearrowright)+ 243 [(\hat d_2 \cdot \hat d_3)^2 + \circlearrowright]\notag \\ 
  &-36 (\hat d_1 \cdot \hat d_3 \hat d_2 \cdot d_3 + \circlearrowright)- 378 (\hat d_1 \cdot \hat d_2 \hat d_2 \cdot \hat d_3 \hat d_3 \cdot \hat d_1) + 243 (\hat d_1 \cdot (\hat d_2 \times \hat d_3))^2\Big ].
\end{align}
These formulas give us all the required ingredients to calculate anomalous Hall conductance. We next proceed to solve the kinetic equation in the exemplary case of $S = 1$. 
\label{eq:AllYouNeedToKnowForSkewScattering}
\end{subequations}

\begin{figure}
    \centering
    \includegraphics[width = .85\textwidth]{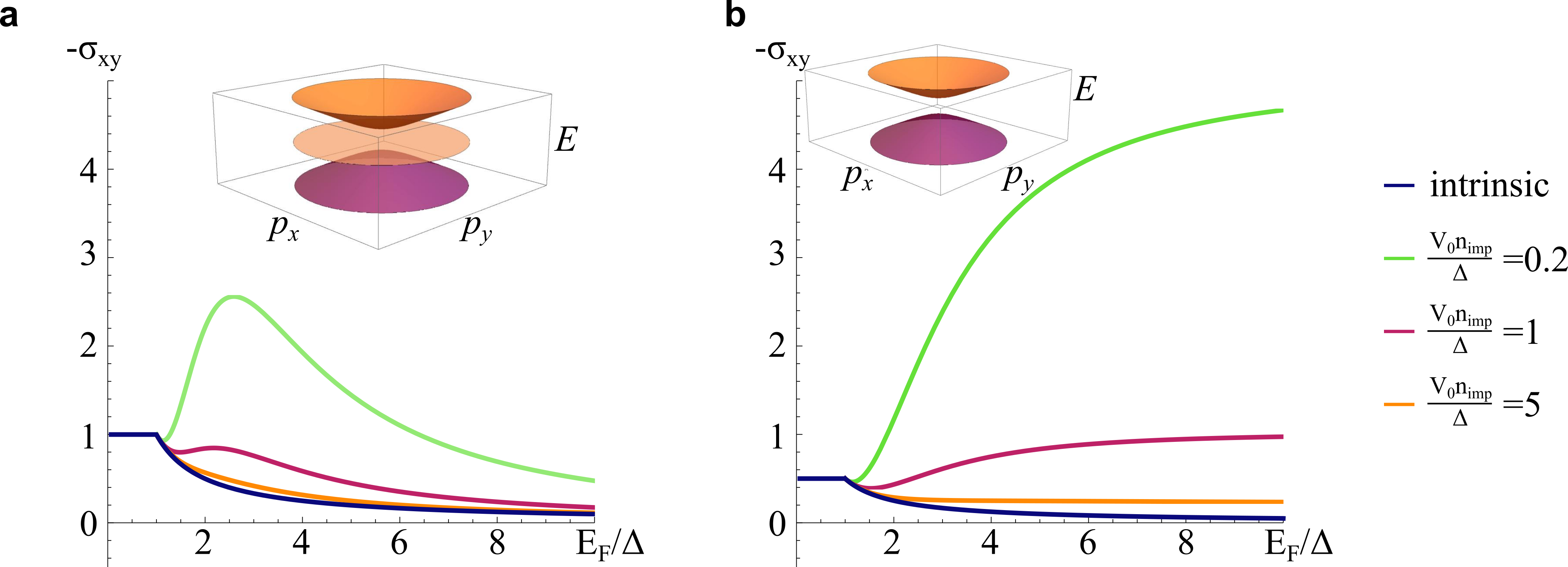}
    \caption{\textbf{a} Anomalous Hall conductivity of massive spin-1 fermions plotted from  Eq.~\eqref{eq:Spin1SigmaXY}. The inset shows the low-energy band structure dispersion of a single massive pseudospin-1 fermion. \textbf{b} Anomalous Hall effect of massive 2D Dirac fermions, Eq.~\eqref{eq:SpinHalfSigmaXY}. The inset shows dispersion of the model.}
    \label{fig:sigmaxyplot}
\end{figure}

\subsection{AHE of pseudospin-1 fermions}
\label{sec:MultifoldAnomalous}

To illustrate the findings of the previous section, we calculate the anomalous Hall response from Eq.~\eqref{eq:Multifold} for $S = 1$, $d_0(\v p) = 0$ and $\v d(\v p) = (v p_x, v p_y, \Delta)$ in a two dimensional system with a single spin-1 touching point. This can be  modelled on a Kagome lattice with an appropriately tuned flux pattern ($\Delta \neq 0$ when the fluxes are not in $\pi \mathbb Z$), see Ref.~\cite{GreenChamon2010} for details. It is worthwhile to highlight that a single spin-1 node may appear in the Brillouin zone without defying the fermion-doubling theorem.

We concentrate on $m = +1$ and take $f = f_{0} + \v p \cdot \v g$, so that the collision integral is
\begin{equation}\label{eq:I-MFF}
    \mathrm{St}_{\text{Born}}\{f\}+\mathrm{St}_{\text{sk}}\{f\} =  - \frac{\v p \cdot \v g  }{\tau} - \frac{\hat e_z \cdot [\v p \times \v g]  }{\tau_{\rm sk}},
\end{equation}
where 
\begin{equation}
    \frac{1}{\tau} = 2 \pi n _{\rm imp} V_0^2 \nu \frac{E^4 + 2 \Delta^2 E^2 + 5 \Delta^4}{8E^4}, \quad \frac{1}{\tau_{\rm sk}} = (2 \pi)^2 n _{\rm imp} V_0^3 \nu^2 \frac{ \Delta^3(E^2 - \Delta^2)}{4E^5} .
\label{eq:TausSpin1}
\end{equation}
Here we used Eqs.~\eqref{eq:overlap-3}, \eqref{eq:Z-2} in the evaluation of Eq.~\eqref{eq:BoltzmannSmallFreq} and assumed point like impurities of strength $V_0$ (for details see \ref{app:Multifold}), the density of states is denoted $\nu = \nu(E) = \theta(E-\Delta) E/(2\pi v^2)$. The solution of the Boltzmann equation to leading order in external static field leads to 
\begin{eqnarray}
    \v g = \frac{v^2}{E_\v p} \left ( \begin{array}{cc}
        1 & - \tau/\tau_{\rm sk} \\
        \tau/\tau_{\rm sk} & 1
    \end{array}\right) \tau e \boldsymbol{\mathcal{E}} \delta(E_F - E_\v p),
\end{eqnarray}
so that $\sigma_{xy} = \sigma_{xy}^{\rm int} + \sigma_{xy}^{\rm sk}$ with respective terms  
\begin{align}
    \sigma_{xy}^{\rm int} &= -e^2 \int_{\v p} f_{0} \Omega_z =  -\frac{e^2}{2\pi\hbar} \frac{\Delta}{E_F}, \notag \\
    \sigma_{xy}^{\rm sk} &= -e ^2\int_{\v p}  \delta(E_{\v p} - E_F) \frac{v^4 p^2}{E_{\v p}^2} \frac{\tau^2}{2\tau_{\rm sk}} = - \frac{e^2}{\hbar} (\Delta\tau) (\nu V_0)\frac{\Delta^2(E_F^2-\Delta^2)^2}{E_F^2(E_F^4 + 2 \Delta^2 E_F^2 + 5 \Delta^4)}.\label{eq:Spin1SigmaXY}
\end{align}
In the final result we restored Planck's constant $\hbar = h/2 \pi$. The relative importance of the side-jump contribution can be also estimated $\sigma^{\text{sk}}_{xy}/\sigma^{\text{sj}}_{xy}\sim (\nu V_0)(E_F\tau)$. For moderately strong impurity potential when, $\nu V_0 \sim 1$, skew scattering dominates in the metallic regime $E_F\tau\gg1$. Note that the flat $m = 0$ band does not contribute, since it's Berry curvature vanishes and it moreover does not intersect the Fermi surface.

Perhaps most notably, the skew scattering contribution to the anomalous Hall response of pseudospin-1 fermions, Fig.~\ref{fig:sigmaxyplot} \textbf{a}, results in a non-monotonic behavior and decays as $\Delta^2/E_F^2$ at large Fermi energies.
This should be contrasted to the more familiar behavior of ordinary spin-1/2 fermions~\cite{SinitsynSinova2007}, governed by Eq.~\eqref{eq:Multifold} for $S = 1/2$ and $d_0(\v p) = 0$ and $\v d(\v p) = 2 (v p_x, v p_y, \Delta)$ (this model has the same density of states $\nu = \nu(E)  = \theta(E-\Delta) E/(2\pi v^2)$ as the spin-1 model). In that case we thereby have 
\begin{align}\label{eq:TausSpinHalf}
    \frac{1}{\tau} = 2\pi n_{\rm imp} V_0^2 \nu \frac{E^2 + 3 \Delta^2}{4E^2},\qquad 
    \frac{1}{\tau_{\rm sk}} = (2\pi)^2 n_{\rm imp} V_0^3 \nu^2 \frac{\Delta(E^2 - \Delta^2)}{8E^3},
\end{align}
so that respective transverse conductivity contributions are given by 
\begin{align}\label{eq:SpinHalfSigmaXY}
    \sigma_{xy}^{\rm int} = - \frac{e^2}{2\pi\hbar} \frac{\Delta}{2 E_F},\qquad
    \sigma_{xy}^{\rm sk} = - \frac{e^2}{\hbar} (\Delta\tau)(\nu V_0) \frac{(E_F^2 - \Delta^2)^2}{4 E_F^2(E_F^2 + 3 \Delta^2)}.
\end{align}
Note that, in contrast to Eq.~\eqref{eq:Spin1SigmaXY}, here the skew scattering contribution saturates at large energies. This crucial distinction of multifold fermions as compared to ordinary spin-1/2 fermions, which is further exemplified in Fig.~\ref{fig:sigmaxyplot} {\bf b}, is a consequence of the differences in the Pancharatnam phase, Eqs.~\eqref{eq:PP}, and ultimately a consequence of a different behavior of wave function overlaps.

\section{Summary and Outlook}\label{sec:Summary}

To conclude, we have presented a systematic derivation of the Boltzmann equation for materials with finite Berry curvature in the regime of small frequency, Eq.~\eqref{eq:BoltzmannSmallFreq}, and large frequency, Eq.~\eqref{eq:BoltzmannLargeFreq}. Our derivation is valid up to second order in driving fields and included a careful treatment of quantum scattering events, such as side jump, skew scattering and shift current contributions. Our results thus serve as the foundation for the theoretical description of linear and nonlinear transport and optics phenomena in topological quantum materials. As an application of our theory, we have considered the anomalous Hall response in multifold fermion systems. Specifically, we have derived the formulae for skew scattering probability, Eqs.~\eqref{eq:LowFreqI03}, \eqref{eq:AllYouNeedToKnowForSkewScattering} which determine the skew scattering for spin-1 and spin-3/2 multifold fermions. As a concrete illustration, we derived the anomalous Hall response for a simple, isotropic model of gapped spin-1 fermions in two dimensions, Eq.~\eqref{eq:Spin1SigmaXY}, see Fig.~\ref{fig:sigmaxyplot}.

Directions of future research include further studies of inelastic skew scattering and side jump contributions due to electron-electron interaction \cite{Pesin2018} in the hydrodynamic regime or electron-phonon interaction \cite{XiaoNiu2019b}. It should be possible to readily incorporated those using a similar diagrammatic technique as exposed in Fig.~\ref{fig:Sigmas}. Moreover, a careful study of the interplay of mixed contributions of photon and impurity scattering for a concrete model is left for the future as well as the generalization of the Boltzmann-Berry equation to non-Abelian Berry curvature. Finally, in regards of multifold fermion systems, which are at the center of the attention for the quantized circular photogalvanic effect, a careful investigation of photocurrent relaxation is of experimental and theoretical interest and readily achievable within the presented formalism.

\section{Acknowledgments}

We thank M. Dzero, S. Li, P. Ostrovsky, D. Pesin, H.-Y. Xie for fruitful discussions and prior collaboration on related topics that inspired this study. We are also grateful to Zongzheng Du for communicating to us on the progress and remaining open question with the diagrammatic calculations of the nonlinear Hall effect. The work of A.L. at UW-Madison was supported by the National Science Foundation CAREER Grant No. DMR-1653661.

\appendix

\section{Collision integrals and averages}
\label{app:Multifold}

In this section of the appendix we present supplemental details regarding calculations carried out in Sec.~\ref{sec:MultifoldAnomalous}. For the point like impurities of strength $V_0$ and the relevant collision integrals that determine intrinsic and skew scattering contributions are given by 
\begin{eqnarray}
    \mathrm{St}^{[2]}_{\text{Born}}\{f\} &=& - 2 \pi n_{\rm imp} V_0^2 \int_{\v p'}  \delta(E_{\v p} - E_{\v p'}) \vert \braket{u_{m, \v p} \vert u_{m, \v p'}} \vert^2 [f(\v p) - f (\v p')], \\
    \mathrm{St}^{[3a]}_{\text{sk}}\{f\} &=& -2 \pi n_{\rm imp} V_0^3 \int_{\v p'\v p''}  \delta(E_{\v p} - E_{\v p'})\delta(E_{\v p} - E_{\v p''})Z_{\v p\v p'\v p''} f(\v p').
\end{eqnarray}
In order to extract scattering times $\tau$ and $\tau_{\text{sk}}$ entering Eq. \eqref{eq:I-MFF} we need to perform several averages over the Fermi surface. These averages can be split in groups of scalars 
\begin{align}
\langle \hat d \cdot \hat d' \rangle_{\hat p'} = \frac{\Delta^2}{E^2}, \qquad \langle (\hat d \cdot \hat d')^2 \rangle_{\hat p'}= \frac{E^4 - 2E^2 \Delta^2 + 3 \Delta^4}{2E^4},
\end{align}
vectors 
\begin{subequations}
\begin{align}
\langle \v p' (\hat d \cdot \hat d') \rangle_{\hat p'} = \frac{E^2 - \Delta^2}{2E^2} \v p,  \qquad
\langle \v p' (\hat d \cdot \hat d')^2 \rangle_{\hat p'} = \frac{\Delta^2(E^2 - \Delta^2)}{E^4} \v p, 
\end{align}
and tensors
\begin{align}
    \langle \v p' \cdot \v g \hat d \cdot (\hat d' \times \hat d'') \rangle_{\hat p', \hat p''} &= \frac{\Delta (E^2 - \Delta^2)}{2 E^3} \hat e_z \cdot (\v p \times \v g), \\
    \langle \v p' \cdot \v g \hat d \cdot (\hat d' \times \hat d'') (\hat d \cdot \hat d') \rangle_{\hat p', \hat p''} &= \frac{\Delta^2}{E^2}\langle \v p' \cdot \v g \hat d \cdot (\hat d' \times \hat d'') \rangle_{\hat p', \hat p''}, \\
   \langle \v p' \cdot \v g \hat d \cdot (\hat d' \times \hat d'') (\hat d' \cdot \hat d'') \rangle_{\hat p', \hat p''} &= \frac{3\Delta^2 -E^2}{2E^2}\langle \v p' \cdot \v g \hat d \cdot (\hat d' \times \hat d'') \rangle_{\hat p', \hat p''},\\
    \langle \v p' \cdot \v g \hat d \cdot (\hat d' \times \hat d'') (\hat d \cdot \hat d'') \rangle_{\hat p', \hat p''} &= \frac{3\Delta^2 -E^2}{2E^2}\langle \v p' \cdot \v g \hat d \cdot (\hat d' \times \hat d'') \rangle_{\hat p', \hat p''}.
\end{align}
\end{subequations}


\bibliography{Berry-BKE}



\end{document}